\documentclass{article}%
\usepackage{xcolor}
\usepackage{amsfonts}
\usepackage{amsmath}%
\usepackage{amssymb}%
\usepackage{graphicx}
\usepackage[square,numbers]{natbib}
\providecommand{\U}[1]{\protect\rule{.1in}{.1in}}
\begin{document}

\title{Sharp values for all dynamical variables via Anti-Wick Quantization\vspace{1cm}\\
\small{
Simon Friederich\\
s.m.friederich@rug.nl\\
University of Groningen, University College Groningen, Hoendiepskade 23/24, 9718BG Groningen, the Netherlands}}

\maketitle

\begin{abstract}This paper proposes an approach to interpreting quantum expectation values that may help address the quantum measurement problem. Quantum expectation values are usually calculated via Hilbert space inner products and, thereby, differently from expectation values in classical mechanics, which are weighted phase-space integrals. It is shown that, by using Anti-Wick quantization to associate dynamical variables with self-adjoint linear operators, quantum expectation values can be interpreted as genuine weighted averages over phase space, paralleling their classical counterparts.

This interpretation arises naturally in the Segal–Bargmann space, where creation and annihilation operators act as simple multiplication and differentiation operators. In this setting, the Husimi Q-function -- the coherent-state representation of the quantum state -- can be seen as a true probability density in phase space. Unlike Bohmian mechanics, the present approach retains the standard correspondence between dynamical variables and self-adjoint operators while paving the way for a classical-like probabilistic interpretation of quantum statistics.
\end{abstract}

\section{Introduction}

Quantum theory would not suffer from the measurement problem if it were unproblematic to combine its empirical success with the assumption that all dynamical variables have sharp -- i.e. non-fuzzy, definite -- values at all times. It allows us to compute probabilities for the possible values of dynamical variables, but -- according to the textbook ``orthodox view'' \citep[p.\ 3]{griffiths} -- only for variables that are ``measured'' or ``observed.'' According to this view, statements about unmeasured dynamical variables are regarded as meaningless. This is profoundly unsatisfying: As \citet{bell_measurement} pointed out, concepts such as ``measurement'' and ``observation'' are vague and anthropocentric. They should not be used to characterize what is real in fundamental physics.

Classical statistical mechanics has no measurement problem: It is not only compatible with, but actually, \textit{presupposes} that all dynamical variables have sharp values at all times. Dynamical variables in classical statistical mechanics are functions $A(\mathbf x,\mathbf p)$ on phase space, the $2N$-dimensional space of canonical coordinates $\mathbf x$ and momenta $\mathbf p$. The probability distribution $P(\mathbf x,\mathbf p)$ expresses (limited) information about the actual location of the system in phase space. The, comparatively, unproblematic character of classical statistical mechanics is reflected in the fact that expectation values of dynamical variables are obtained as phase space integrals of the dynamical variables $A(\mathbf x,\mathbf p)$, weighted by the probability distribution $P(\mathbf x,\mathbf p)$:
\begin{eqnarray}
\langle A \rangle = \int_{\mathbb R^{2N}} A(\mathbf x,\mathbf p)\,P(\mathbf x,\mathbf p)\,d\mathbf x\,d\mathbf p\,. \label{classical_expectation}
\end{eqnarray}

In quantum theory, in contrast, the role of the probability distribution $P(\mathbf x,\mathbf p)$ is taken over by the quantum state $|\psi\rangle$, which is a unit vector in Hilbert space, defined up to a global phase factor,  or, more generally, by the density matrix $\hat\rho$. Dynamical variables are promoted to Hilbert space self-adjoint (or, more generally, essentially self-adjoint \citep[Ch.\ 9]{hall}) operators $\hat A$, and expectation values are computed as Hilbert space inner products:
\begin{eqnarray}
\langle A \rangle = \langle \psi| \hat A|\psi\rangle\,. \label{quantum_expectation}
\end{eqnarray}
According to conventional wisdom, Eq.\ (\ref{quantum_expectation}) cannot be interpreted as a phase space integral along the lines of Eq.\ (\ref{classical_expectation}). In fact, a large number of rigorous mathematical results are widely taken to indicate that assuming a definite phase space location for a system or, equivalently, assining sharp, non-fuzzy, values to all its dynamical variables would have an unacceptably high price. Such an assignment would, notably, have to be \textit{contextual} according to the Kochen-Specker theorem \citep{kochenspecker}  -- making such assignments appear ad hoc and unnatural -- it would have to be  \textit{non-local} according to Bell's theorem \citep{belltheorem}  -- threatening conflict with relativity theory -- and, according to the PBR theorem \citep{pbr}, would have to ascribe reality to the quantum state, in contrast to classical statistical mechanics where the probability distribution has a much more organic, broadly speaking, \textit{epistemic} interpretation (but see \citep{myrvold} for nuance).

The transition from a classical to a quantum theory can be associated with the procedure of \textit{quantization}: mapping dynamical variables qua functions on phase space to self-adjoint operators on a Hilbert space, subject to certain constraints and desiderata. There is a vast literature in mathematical physics that compares the respective advantages and disadvantages of different quantization schemes for a variety of spaces on which quantum theories may be defined. (See \citep{folland,shubin,gosson,ali,landsman} for introductions and reviews of important results.)

However, somewhat surprisingly in view of the multitude of quantization schemes that have been proposed over the decades, there does not seem to have been much interest in whether the difficulties of interpreting quantum expectation values computed as Hilbert space inner products according to Eq.\ (\ref{quantum_expectation}) could perhaps be avoided via a judicious choice of quantization scheme. The idea would be that, for a suitable quantization mapping $A\mapsto\hat A$, the Hilbert space inner products $ \langle \psi| \hat A|\psi\rangle$ are identical with weighted phase space averages $ \int_{\mathbb R^{2N}} A(\mathbf x,\mathbf p)\,P(\mathbf x,\mathbf p)\,d\mathbf x\,d\mathbf p$ with suitably chosen probability distributions $P(\mathbf x,\mathbf p)$.  The present paper aims to fill this lacuna by exploring the idea that the measurement problem (or, more specifically, the difficulties with assining sharp values to all dynamical variables) might at least partially be an artifact of an infelicitious choice of quantization scheme.

Our investigation will confirm this suspicion: If dynamical variables are promoted to self-adjoint operators via \textit{Anti-Wick} quantization -- an independently attractive quantization scheme that exemplifies a versatile and widely applicable approach to quantization championed by \citet{berezin} -- their quantum expectation values computed via Eq.\ (\ref{quantum_expectation}) have a straightforward interpretation as classical phase space integrals of the form Eq.\ (\ref{classical_expectation}). One may interpret this finding as suggesting that, when it comes to solving the measurement problem, Anti-Wick quantization is more promising than the more well-known Weyl quantization. If Weyl quantization is used to connect dynamical variables and self-adjoint operators, then, as will be shown, the expectation values of many observables $A$ can still be calculated as weighted phase space integrals, but, weirdly, with the observable $A$ being replaced in the integral by the ``Anti-Wick symbol'' of its Weyl operator. The present finding turns out to be in line with -- and provides additional motivation for -- an interpretation of quantum theory gestured at originally by Fritz Bopp \citep{bopp} and more recently developed by Drummond and Reid as well as Friederich \citep{drummondreid,friederich} that regards the Husimi Q-function as a proper probability distribution on phase space.

Anti-Wick quantization is closely related to the \textit{Segal-Bargmann space}, a version of quantum Hilbert space where state vectors correspond to holomorphic functions  and creation and annihilation operators act as simple multiplication and differentiation operators. While the technical elements of Anti-Wick quantization and Segal-Bargmann space are well known in mathematical physics and quantum optics, they are hardly ever considered in the foundations literature. One of the aims of this paper is therefore to draw foundationalists'attention to Segal-Bargmann space, how it provides a natural home for Anti-Wick quantization allows to interpret quantum expectation values as genuine phase-space averages.

\section{From phase space to Hilbert space}

There is one straightforward way in which one can read any ``classical'' expectation value $\int_{\mathbb R^{2N}} A(\mathbf x,\mathbf p)\,P(\mathbf x,\mathbf p)\,d\mathbf x\,d\mathbf p$ as a Hilbert space scalar product  $\langle f| \hat A|f\rangle$, namely, by choosing as the Hilbert space the space $\mathcal L^2(\mathbb R^{2N})$ of square-integrable functions, selecting some $f\in\mathcal L^2(\mathbb R^{2N})$ for which $P(\mathbf x,\mathbf p)=|f(\mathbf x,\mathbf p)|^2$ everywhere and defining $\hat A|f\rangle$ via pointwise multiplication of $A$ and $f$. But a little reflection shows that this Hilbert space and these operators cannot be the ``right ones'' with which we are all familiar from quantum theory: Pointwise multiplication is commutative, so the operators representing, for instance, position $\mathbf x$ and momentum $\mathbf p$ on this Hilbert space are clearly not the familiar non-commuting ones.

This observation should not discourage us from our attempt to seek an interpretation of quantum expectation values $\langle \psi| \hat A|\psi\rangle$ as weighted phase space averages. A scenario that we may want to reckon with, to be elaborated a bit more in Section 5, is that the (at present unknown) dynamics of phase space locations $\mathbf z$ may not give rise to a law of time evolution for the probability distributions $P(\mathbf x,\mathbf q)$ that is as practical to work with as the Liouville equation for the phase space probability density in classical statistical mechanics. However, it may be the case that, for a more restricted \emph{class} of probability densities $P(\mathbf x,\mathbf q)$ their time evolution can be captured efficiently via expressing these probability densities through quantum states and their time evolution via the Schr\"odinger equation.

Indeed there is a restricted class of phase space probability densities which seems promising in this regard in that it they can be expressed in terms of phase space functions $F$  forming a Hilbert space so that calculating the expectation value of a dynamical variable $A$ can be expressed in the familiar way Eq.\ (\ref{quantum_expectation}) from quantum mechanics. This is the so-called \emph{Segal-Bargmann space}, which will play a crucial role in the argument of this paper. To define it, one must express the phase space coordinates $x_j$ and $p_j$ through dimensionless complex variables $z_j$ via
\begin{eqnarray}
z_j=\frac{1}{\ell}x_j-i \frac{\ell}{\hbar}p_j\,, \nonumber\\
\overline{z_j}= \frac{1}{\ell}x_j+i \frac{\ell}{\hbar}p_j\,,\label{complex_variable}
\end{eqnarray}
where $\ell$ is some suitably chosen length scale, for instance the Compton length, or $\ell=\hbar/(m\omega)$ in the case of the harmonic oscillator. In what follows, we put $\ell=1$ for simplicity. The reader may wonder why $z_j$ is defined as $\frac{1}{\ell}x_j-i \frac{\ell}{\hbar}p_j$ rather than $\frac{1}{\ell}x_j+i \frac{\ell}{\hbar}p_j$, which is the more common convention for defining the complex phase space variable. The answer is that the convention used here is the standard one for Segal-Bargmann space (see \citep[Eq.\ (13.1)]{hall}), which allows creation and annihilation operators to be given by multiplication and differentiation with respect to the complex phase space variable. Results obtained using standardly defined phase space variables $\alpha_i=\frac{1}{\ell}x_j+i \frac{\ell}{\hbar}p_j$ can be recovered by using $\boldsymbol \alpha=\overline{\mathbf z}$.

Segal-Bargmann space, commonly denoted $\mathcal H L^2(\mathbb C^n,\mu_\hbar)$, is the space of  holomorphic functions $F$ defined on $\mathbb C^n$ whose modulus squared can be integrated when weighted with a Gaussian of width $\hbar$ (which at this point is treated as an adjustable constant $>0$):
\begin{eqnarray}
\int_{\mathbb C^n}|F(\mathbf z)|^2\mu_\hbar(\mathbf z)\,d\mathbf z\,<\,\infty\,,
\end{eqnarray}
where $\mu_\hbar$ is the Gaussian
\begin{eqnarray}
\mu_\hbar(\mathbf z)=\frac{1}{(\pi\hbar)^n}e^{-|\mathbf z|^2/\hbar}\,.
\end{eqnarray}

The inner product on Segal-Bargmann space is defined by:
\begin{eqnarray}
\langle F,G\rangle=\int_{\mathbb C^n}\overline{F(\mathbf z)}G(\mathbf z)\mu_\hbar(\mathbf z)\,d\mathbf z\,.\label{inner_product}
\end{eqnarray}

It can be shown (see \cite[Prop. 14.15]{hall}) that the Segal-Bargmann space is dense with respect to the norm
\begin{eqnarray}
||F||=\sqrt{\langle F,F\rangle} \label{norm}
\end{eqnarray}
and a Hilbert space with respect to the inner product Eq.\ (\ref{inner_product}).

As announced above, from the perspective of the present analysis, where we seek to somehow recover quantum states as encoding probability densities on phase space, focusing on Segal-Bargmann space\footnote{The considerations presented here can straightforwardly be generalized to probability distributions that are weighted sums of probability distributions of the form Eq.\ (\ref{SB_form}). This generalization corresponds to widening the scope from considering vectors $|F\rangle$ on Segal-Bargmann space to density matrices. The focus on pure states is chosen merely to make Segal-Bargmann space more accessible to readers who are yet unfamiliar with it.} effectively means restricting consideration to phase space probability distributions $P(\mathbf z)$ which can be expressed via vectors $F$ in Segal-Bargmann space as
\begin{eqnarray}
P(\mathbf z)=|F(\mathbf z)|^2\mu_\hbar(\mathbf z)\,. \label{SB_form}
\end{eqnarray}
Due to the definition Eq.\ (\ref{norm}) of the norm on Segal-Bargmann space $P(\mathbf z)$ is normalized to $1$ if and only if $F$ is a unit vector in Segal-Bargmann space.

It is helpful to consider which probability distributions on phase space can, and which ones cannot, be written in the form  of Eq. (\ref{SB_form}). Notably, probability distributions with discontinuities such as the characteristic functions $\chi_\Delta$ of specific phase space regions $\Delta$ cannot be written in this form due to the requirement that $F$ be holomorphic. Moreover, as will be discussed later, only probability distributions with variances above a certain minimal (in terms of $\hbar$) threshold in any coordinate direction on phase space can be written in the form  Eq. (\ref{SB_form}). That means, while the phase space location $\mathbf z$ will be treated as ``sharp'' and real in this paper, the probability densities considered here are not ``sharp'' in the sense that they are all spread out to a certain degree. For the time being, one may regard the restriction of consideration to probability distributions which can be written as Eq. (\ref{SB_form}) as a version of Spekkens's information-theoretic constraint that restricts how much knowlegde one can have, at any given time, about the ontic state of a system in his famous toy theory \citep{spekkens_toy}.\footnote{It is interesting to note -- and perhaps worthwhile to explore in future work -- that Spekkens imposes his information-theoretic constraint in a model that he uses to motivate the epistemic view of quantum states. This motivation fits nicely with the present work, which is likewise naturally combined with a (broadly speaking) epistemic interpretation of quantum states and the probability distributions that they encode as analogous to the probability distributions in classical statistical mechanics, which are usually seen as non-ontic.} Ultimately, any such restriction will need a thorough and rigorous justification. In Section 5 we will gesture at the direction from which we expect that justification to come (namely to reveal the restriction to be an indirect consequence of, at present hypothetical, underlying microdynamics), but a more in-depth discussion and justification will have to be given elsewhere.

\section{Weighted phase space integrals as Segal-\-Barg\-mann space inner products}

In this section, we will show that, for probability distributions that can be written in the form Eq.\ (\ref{SB_form}), expectation values of dynamical variables $A(\mathbf x,\mathbf p)$ computed as phase space integrals in the form Eq.\ (\ref{classical_expectation}) can be rewritten using Anti-Wick-quantized self-adjoint operators $\hat A$ on Segal-Bargmann space as inner products in the form Eq.\ (\ref{quantum_expectation}).

The expectation value of a dynamical variable $A(\mathbf z)$, computed as a classical phase space integral over $\mathbf z$ rather than $\mathbf x$ and $\mathbf p$, is
\begin{eqnarray}
\langle A\rangle=\int_{\mathbb C^n}A(\mathbf z)\,P(\mathbf z)\,d\mathbf z\,.\label{classical_expectation_new}
\end{eqnarray}
If $P(\mathbf z)$ can be written in the form  of Eq. (\ref{SB_form}), this can be re-expressed as
\begin{eqnarray}
\langle A\rangle&=&\int_{\mathbb C^n}A(\mathbf z)\,|F(\mathbf z)|^2\mu_\hbar(\mathbf z)\,d\mathbf z\nonumber\\
&=&\int_{\mathbb C^n}\overline{F(\mathbf z)}A(\mathbf z)\,F(z)\,\mu_\hbar(\mathbf z)\,d\mathbf z\,.\label{final_integral}
\end{eqnarray}
We now use that the \textit{coherent state vector} $|\chi_{\mathbf z}\rangle$ defined by the function
\begin{eqnarray}
\chi_{\mathbf z}(\mathbf w)=e^{\overline{\mathbf z}\cdot \mathbf w/\hbar}\,,\label{coherent_state}
\end{eqnarray}
is a so-called \textit{reproducing kernel} of Segal-Bargmann space. It has the distinctive ``reproducing'' property that, for any $F\in \mathcal H L^2(\mathbb C^n,\mu_\hbar)$,
\begin{eqnarray}
F(\mathbf z)=\langle \chi_{\mathbf z}|F  \rangle=\int_{\mathbb C^n}e^{\mathbf z\cdot \overline{\mathbf w}/\hbar} F(\mathbf w)\mu_\hbar(\mathbf w)\,d\mathbf w\,.\label{kernel}
\end{eqnarray}
(See \citep[Prop.\ 14.17]{hall} for this statement together with a proof.)

This property allows one to rewrite the last line of Eq.\ (\ref{final_integral}), turning it into an inner product on Segal-Bargmann space:
\begin{eqnarray}
\langle A\rangle&=&\int_{\mathbb C^n}A(\mathbf z)\,\overline{F(\mathbf z)}\,F(\mathbf z)\,\mu_\hbar(\mathbf z)\,d\mathbf z\nonumber\\
&=&\int_{\mathbb C^n}A(\mathbf z)\,\langle F|\chi_{\mathbf z}  \rangle\,\langle \chi_{\mathbf z}|F  \rangle\,\mu_\hbar(\mathbf z)\,d\mathbf z\nonumber\\
&=&\langle F|\,\left(\int_{\mathbb C^n}A(\mathbf z) |\chi_{\mathbf z}  \rangle\,\langle \chi_{\mathbf z}|\mu_\hbar(\mathbf z)d\mathbf z\right)\,|F\rangle\nonumber\\
&\equiv&\langle F|\hat A| F  \rangle \label{main_result}
\,,\label{rewrite}
\end{eqnarray}
where the Hilbert space operator $\hat A$ has been introduced, which is defined by how it acts on a vector $|F\rangle$ in Segal-Bargmann space:
\begin{eqnarray}
\hat A |F\rangle=\int_{\mathbb C^n}A(\mathbf z)|\chi_{\mathbf z}  \rangle\,\langle \chi_{\mathbf z}|F\rangle\mu_\hbar(\mathbf z)\,d\mathbf z\,.\label{AW_define}
\end{eqnarray}
The operator $\hat A$, so defined, is sometimes referred to as a \textit{Toeplitz operator}. It has the attractive property that $\hat A| F\rangle$ is a vector in Segal-Bargmann space (and, a fortiori, holomorphic), even if the dynamical variable $A(\mathbf z)$ is not holomorphic. In other words, the operator $\hat A$ turns multiplication with (in general non-holomorphic) $A(\mathbf z)$ into an operation that preserves being holomorphic but allows one to compute the same expectation values as multiplication with $A$.

Now, the mapping of dynamical variables $A$ onto self-adjoint operators $\hat A$ encoded in Eq. (\ref{AW_define}) coincides exactly with the definition of Anti-Wick quantization on Segal-\-Barg\-mann space, see \citep[Eq.\ (2.93)]{folland}. Eq.\ (\ref{main_result}), in turn, contains the central observation of this paper: Expectation values $\langle A\rangle$, defined via phase space integrals, for probability distributions conforming to Eq.\ (\ref{SB_form}), can be re-expressed as Hilbert space inner products, provided that the phase space dynamical variable $A$ is mapped to the self-adjoint operator assigned to it by Anti-Wick quantization.

It is instructive to consider the form that the derivation of Eq.\ (\ref{main_result}) and the form of the Anti-Wick operator Eq.\ (\ref{AW_define}) take for polynomial observables specifically. Focusing on the case of one spatial dimension for simplicity, where polynomial observables have the form $A(z)=\sum_{j,k}A_{jk}z^j\overline z^k$, the expectation value becomes:
\begin{eqnarray}
\langle A\rangle&=&\int_{\mathbb C^n}\sum_{j,k}A_{jk}z^j\overline z^k\,|F(z)|^2\mu_\hbar(z)\,dz\nonumber\\
&=&\int_{\mathbb C^n}\sum_{j,k}A_{jk}\overline z^k\,\overline{F(z)}z^j\,F(z)\,\mu_\hbar(z)\,dz\\
&=&\int_{\mathbb C^n}\overline{F(z)}\sum_{j,k}A_{jk}\overline z^k z^j\,F(z)\,\mu_\hbar(z)\,dz\,.\nonumber
\end{eqnarray}
Using integration by parts and Lemmas 14.12 and 14.13 in \citep{hall}, this can be rewritten as
\begin{eqnarray}
\langle A\rangle&=&\int_{\mathbb C^n}\overline{F(z)}\sum_{j,k}A_{jk}\frac{\partial}{\partial z^k}z^j\,F(z)\,\mu_\hbar(z)\,dz\,\label{polynomial_integral}
\end{eqnarray}
It is customary to define creation and annihilation operators $a^\dagger_j$ and $a_j$ on the space of holomorphic polynomials in Segal-Bargmann space via 
\begin{eqnarray}
a^\dagger_j&=&z_j\\
a_j&=&\hbar\frac{\partial}{\partial z_j}\,.
\end{eqnarray}
So defined, these operators fulfil the (apart from a convention-dependent appearance of $\hbar$) usual commutation relations for creation and annihilation operators:
\begin{eqnarray}
\lbrack a_j,a_k\rbrack&=&0\,,\nonumber\\
\lbrack a^\dagger_j,a^\dagger_k\rbrack&=&0\nonumber\\
\lbrack a_j,a^\dagger_k\rbrack&=&\hbar\delta_{j,k}\,. \label{commutation_relation}
\end{eqnarray}
Segal-Bargmann space with its inner product Eq.\ (\ref{inner_product}) is defined precisely such that these operators are adjoints of each other.

Using the so-defined creation and annihilation operators, which in the one-dimensional case are simply $a^\dagger$ and $a$, the phase space integral in Eq.\ (\ref{polynomial_integral}) can again be expressed as an inner product:
\begin{eqnarray}
\langle A\rangle&=&\langle F,\,\sum_{j,k}A_{jk}a^j(a^\dagger)^k\,F\rangle. \label{expectation_as_inner_product}
\end{eqnarray}
In other words, for polynomial observables $A(z)=\sum_{j,k}A_{jk}z^j\overline z^k$, their expectation values $\langle A\rangle$, defined via phase space integrals as in classical statistical mechanics, can be re-expressed as Hilbert space inner products, provided that the phase space dynamical variable $A$ (assumed to be real) is ``promoted'' to a self-adjoint operator according to
\begin{eqnarray}
\sum_{j,k}A_{jk}z^j\overline z^k\mapsto\sum_{j,k}A_{jk}a^j(a^\dagger)^k\,. \label{Anti-Wick}
\end{eqnarray}
Eq.\ (\ref{Anti-Wick}) is a special case of Anti-Wick quantization as encoded in Eq.\ (\ref{AW_define}).

The order of the creation and annihilation operators in Eq. (\ref{Anti-Wick}), with annihilation operators to the left of creation operators -- also known as ``anti-normal'' or ``anti-Wick'' ordering -- matters crucially. If this order were switched, or if, say, annihilation and creation operators were treated symmetrically when defining the self-adjoint operator $\hat A$, then $\langle A\rangle$, understood as a classical phase space integral, could no longer be calculated as $\langle F,\hat A F\rangle$.

The Stone-von Neumann theorem allows one to recover any expectation values that can be calculated via Segal-Bargmann space in terms of the more well-known Schr\"odinger space $L^2(\mathbb R^n)$ (or, for that matter, any other Hilbert space on which an appropriate representation of position and momentum operators is defined). (See e.g. \citep[Sects. 3.3 and 3.4]{hall} for concise definitions of these operators.) Anti-Wick quantization on Schr\"odinger space, in analogy to Anti-Wick quantization on Segal-Bargmann space (Eq.\ (\ref{AW_define}), is defined by the Toeplitz operator (see \citep[Eq.\ (11.54)]{gosson})
\begin{equation}
\hat{A}=\frac{1}{\pi^n}\int_{\mathbb{C}^{n}}A(\mathbf z)|\mathbf z\rangle\langle \mathbf z|\, d\mathbf z\label{aw6}
\end{equation}
where $|\mathbf z\rangle\langle \mathbf z|$ is the orthogonal projection on the standard coherent state associated with $\mathbf z$ in Schr\"odinger space $L^2(\mathbb R^n)$. (See \citep[Ch.\ 11]{gosson} and \citep[\S 24]{shubin} for more details and definitions.)

The unitary map that connects equivalent vectors in Schr\"odinger and Segal-Bargmann space is called the Segal-Bargmann transform $B$. Taken together, our considerations show that an expectation value $\langle A\rangle$  defined as a classical phase space integral with a probability distribution $P(\mathbf z)$, can be recovered as a quantum expectation value
\begin{equation}
\langle A\rangle =\langle\psi|\hat A|\psi\rangle
\end{equation}
with a state vector $\psi\in L^2(\mathbb R^n)$ from Schr\"odinger space if
\begin{itemize}
\item $P(\mathbf z)$ can be written in the form $|F(\mathbf z)|^2\mu_\hbar(\mathbf z)$ with $|F\rangle$ a vector in Segal-Bargmann space and $\mu_\hbar(\mathbf z)$ the Segal-Bargmann space measure,
\item $|F\rangle=B|\psi\rangle$ is the Segal-Bargmann transform of $|\psi\rangle$,
\item $\hat A$ is obtained from $A$ via Anti-Wick quantization (Eq.\ (\ref{aw6}))\,.
\end{itemize}

\section{The Q-function and empirical adequacy}

Anti-Wick quantization is only one among several frequently considered and investigated quantization schemes on $\mathbb R^{2m}$ or $\mathbb C^n$, and not the most widely used one. Most often preference is given to Weyl quantization, which promotes polynomials of position and momentum to symmetrized operators. In one spatial dimension this means \citep[Def. 13.1]{hall}
\begin{eqnarray}
Q_{Weyl}(x^jp^k)=\frac{1}{(j+k)!}\sum_{\sigma\in S_{j+k}}\sigma(\hat X, \hat X, ..., \hat X, \hat P, \hat P, ..., \hat P)\,,\label{Weyl_mapping}
\end{eqnarray}
where $\hat X$ is the position operator, $\hat P$ is the momentum operator, and $S_{j+k}$ is the set of all permutations of $j+k$ objects. Yet another frequently considered quantization scheme is Wick quantization, which differs from Anti-Wick quantization in that creation operators are placed to the left of annihilation operators (which is often the more convenient order to consider in quantum field theory).

The difference between these quantization schemes can be illustrated with the dynamical variable $f(x,p)=x^2$, which is promoted to, respectively (see \citep[Ex. 13.2]{hall}),
\begin{eqnarray}
Q_{Weyl}(x^2)&=&\hat X^2\,\nonumber\\
Q_{Wick}(x^2)&=&\hat X^2-\frac{1}{2}\hbar I\,\nonumber\\
Q_{AW}(x^2)&=&\hat X^2+\frac{1}{2}\hbar I\,\label{AW_squared}
\end{eqnarray}
Weyl quantization is frequently regarded as the most attractive quantization scheme because it maximizes the degree to which the commutator can be seen as the quantum analogue of the Poisson bracket $\lbrace f,g\rbrace$. More concretely, if $f$ and $g$ are polynomials in $x$ and $p$ and $f$ is of degree at most $2$, then
\begin{eqnarray}
\frac{1}{i\hbar}\lbrack Q_{Weyl}(f),Q_{Weyl}(g)\rbrack= Q_{Weyl}(\lbrace f,g\rbrace)\,.\label{commutator}
\end{eqnarray}
According to Groenewold's theorem \citep{groenewold}, however, this does not in general hold if $f$ is of degree $3$ or higher (See \citep[Ch.\ 13.4]{hall} for an accessible statement and proof). For all other quantization schemes, the relation Eq.\ (\ref{commutator}) breaks down already for polynomials of lower degrees, making Weyl quantization appear privileged. However, arguably this is at most a very weak reason to prefer Weyl quantization over other quantization schemes, firstly because maximizing the parallel between the commutator and the Poisson bracket seems to be mostly an aesthetic criterion without any independent systematic advantages, and secondly because, as remarked, even in Weyl quantization the parallel remains limited.

The result of the previous section, in contrast, gives a very robust reason for, from a foundational point of view, preferring Anti-Wick quantization, namely, that it allows us to interpret the calculation of expectation values via inner products on Hilbert space straightforwardly as a shorthand for a phase space integrals. This allows us to assume that, at any given instant in time at least, the system has a definite phase space location and, hence, that all dynamical variables have sharp values. For this move to work, the only constraint that we had to impose, to recall, was that the phase space probability distribution can be written in the form  Eq. (\ref{SB_form}), as $|F(\mathbf z)|^2\mu_\hbar(\mathbf z)$, where $|F\rangle$ is an element of Segal-Bargmann space.

In quantum theory, the probability distributions which can be written in this form have in fact long been known, in a different guise, namely, as the so-called Husimi Q-functions \citep{husimi}, which, for pure states $|\psi\rangle$ are defined by
\begin{eqnarray}
Q_{|\psi\rangle}(\mathbf z) = \frac{1}{\pi^n}|\langle\phi_{\mathbf z}|\psi\rangle|^2\,, \label{Qdef}
\end{eqnarray}
where $|\phi_{\mathbf z}\rangle$ is the coherent state centred at $\mathbf z$.

To see why the probability density $P(\mathbf z)$ used in the phase space integrals in the previous section can be identified with the Q-function, we may use the definition Eq.\  (\ref{aw6}) of the Anti-Wick operator $\hat{A}=\frac{1}{\pi^n}\int_{\mathbb{C}^{n}}A(\mathbf z)|\mathbf z\rangle\langle \mathbf z| d\mathbf z$ in Schr\"odinger space $L^2(\mathbb R^n)$. We obtain for the quantum expectation value in Anti-Wick quantization, calculated according to Eq.\ (\ref{quantum_expectation}):
\begin{eqnarray}
\langle A \rangle &=& \langle \psi|\hat A|\psi\rangle\nonumber\\
&=&\langle \psi|\left(\,\frac{1}{\pi^n}\int_{\mathbb{C}^{n}}A(\mathbf z)|\phi_{\mathbf z}\rangle\langle\phi_{\mathbf z}|\, d\mathbf z\right)\,|\psi\rangle \\
&=&\frac{1}{\pi^n}\int_{\mathbb{C}^{n}}A(\mathbf z)\langle\psi|\phi_{\mathbf z}\rangle\langle\phi_{\mathbf z}|\psi \rangle \,d\mathbf z\\
&=&\int_{\mathbb{C}^{n}}A(\mathbf z)\frac{|\langle\phi_{\mathbf z}|\psi\rangle|^2|}{\pi^n} d\mathbf z\,.
\end{eqnarray}
Comparison with Eq.\ (\ref{final_integral}) shows that, after implementing the definition Eq.\ (\ref{Qdef}) of the Husimi Q-function:
\begin{eqnarray}
Q(\mathbf z) = |F(\mathbf z)|^2\mu_\hbar(\mathbf z)\,.
\end{eqnarray}
This establishes that the restriction to probability distributions of the form Eq.\ (\ref{SB_form}) is a restriction to the Husimi functions of pure quantum states. Allowing weighted sums of such probability distributions amounts to including the Husimi functions of mixed states.

The Husimi Q-function, which can also be obtained by a Weierstrass transformation (smoothing with a Gaussian filter involving the characteristic energy scale of the theory) from the more well-known Wigner function \citep{wigner}, is often used among other so-called quasi-probability distributions, including the Wigner function, in a variety of calculations, notably in quantum optics. Embracing Anti-Wick quantization as  encoding how dynamical variables and self-adjoint operators are connected amounts to interpreting the Husimi Q-function as a genuine probability distribution on phase space -- the one that reflects the relative frequencies of phase space locations for systems prepared ``in'' the quantum states corresponding to the associated Segal-Bargmann vectors $F$. Intriguingly, this move -- interpreting the Q-function as a proper phase space probability density in quantum theory -- was considered long ago by Fritz Bopp\footnote{Bopp's 1956 paper, curiously written in French \citep{bopp} focuses on a quantity equivalent to the Husimi function which he explicitly interprets as ``la densit\'e de la probabilit\'e dans l'espace des phases.'' However, neither Anti-Wick quantization nor Segal-Bargmann space were available at that time, and Bopp therefore paired his distribution with a differential-operator calculus now known to correspond to the Weyl-Moyal formalism. The present paper develops the direction that Bopp could not pursue: namely, coupling the Husimi probability density to Anti-Wick quantization, which uniquely restores classical-looking expectation values and brings out the natural role of Segal–Bargmann space.} \cite{bopp} and has recently been reconsidered \citet{drummondreid} and \citet{friederich}, the latter as ``Q-based interpretation.''

It should be emphasized that the present account does not regard the Husimi Q-function as a probability density over coherent states. Rather, it is a probability density over the phase space points themselves, which are taken to be the ontic states of the system. Coherent states simply provide a convenient Hilbert-space representation of those points: they serve as ``addresses'' allowing us to evaluate the density operator at any chosen phase-space location. As quantum states, coherent states are therefore epistemic objects, just as classical probability densities are epistemic in classical statistical mechanics; the only ontic ingredient in the present framework is the actual phase-space point.

The idea of interpreting the Q-function as a proper probability density on phase space is arguably very attractive in the light of the features of Segal-Bargmann space presented in the previous section. The reason why hardly anyone seems to have seriously considered it before may be that it does not reproduce the marginals familiar from Weyl quantization.\footnote{Another reason may be that there are rigorous results due to \citet{ferrieemerson} and \citet{spekkens_negativity} which rule out accounting for the empirically well-confirmed predictions of quantum theory using positive semi-definite quasi-probability distributions. However, these results do not apply to the interpretation of the Husimi Q-function as a proper probability density because, as explained by \citet[Sect.\ 5]{friederich}, this interpretation does not fulfil the assumption of \emph{$\lambda$-mediation} that is used in the proofs of these theorems.} In Weyl quantization, for a pure quantum state $|\psi\rangle$, the probability density at the position coordinate $\mathbf x$ is famously given by $|\langle\mathbf x|\psi\rangle|^2$. According to Anti-Wick quantization, in contrast, that probability density is given by the phase space integral of the $Q$-function over canonical momenta $\mathbf p$:
\begin{eqnarray}
\int Q(\mathbf x, \mathbf p) d\mathbf p \neq |\langle\mathbf x|\psi\rangle|^2\,. \label{empirical_inadequacy}
\end{eqnarray}

In view of the discrepancy between the left and right hand side of Eq.\ (\ref{empirical_inadequacy}), and because quantum theory is usually applied based on the assumption that $|\langle\mathbf x|\psi\rangle|^2$ is the probability density in configuration space, one may think that Anti-Wick quantization could be empirically inadequate in the light of well-established experimental findings. However, while we cannot rule this out completely, we do not believe this to be the case. In measurement contexts where full phase space location is measured, e.g. so-called heterodyne mode detection in quantum optics, it is predictably the Husimi Q-function that provides the correct measurement statistics \citep{PhysRevLett.117.070801,wiseman_milburn}. In general, it specifies the distribution of results in ``retrodictively optimal'' phase space measurements \citep{Appleby1999_OptimalJoint}. In any case, as \citet{bell} noted when discussing Bohmian mechanics, the true bar for empirical adequacy of an interpretation of quantum theory is that it recover the empirically confirmed probabilities and expectation values for variables that describe macroscopically distinct configurations of macro-objects such as measurement devices to which we have more or less direct empirical access. Weyl and Anti-Wick quantization are known to be ``equivalent'' in that they coincide in the limit $\hbar\mapsto 0$ \citep[Sect.\ 4.3]{landsman}, and discrepancies between the left and right hand sides of  Eq.\ (\ref{empirical_inadequacy}) are always of the order $\hbar$ at most, so, depending on the characteristic energy scale used in the Weierstrass transformation which connects the Wigner and Q-functions, these discrepancies may not be detectable for macroscopic ``pointer'' observables. \citet[Sect.\ 5]{friederich} subjects a typical measurement context to scrutiny, where a micro-system is coupled to a macroscopic apparatus, and finds the discrepancies between Weyl and Anti-Wick quantization do indeed not appear at the level of the pointer observables if pointer positions of distinct outcomes are assumed to be macroscopic and discrete.

Note that the conclusion obtained in the previous section that quantum expectation values $\langle A \rangle = \langle \psi| \hat A|\psi\rangle$ can be calculated as phase space integrals applies for many observables $A$ \textit{even if Weyl quantization rather than Anti-Wick quantization is used to obtain $\hat A$ from $A$}. However, in that case the phase space integral is not over the observable $A$ itself, but instead over the observable, if it exists\footnote{Not every Weyl operator $\hat A$ has an Anti-Wick symbol \citep[p.\ 140f.]{folland}.}, which is promoted to $A$'s Weyl operator $\hat A_{Weyl}$ by Anti-Wick quantization, i.e. the ``Anti-Wick symbol'' of $\hat A_{Weyl}$. In other words, regarding Weyl quantization as encoding the connection between dynamical variables and self-adjoint operators commits us to the claim that, for observables $A$ whose Weyl operator $\hat A_{Weyl}$ has an Anti-Wick symbol (which includes, notably, all operators that are polynomials in $P$ and $Q$), the expectation value in the quantum state $|\psi\rangle$ is given as the phase space integral
\begin{eqnarray}
\langle A\rangle = \int Q_{AW}^{-1}\left(Q_{Weyl}(A)\right)(\mathbf z)\,P_\psi(\mathbf z)\,d\mathbf z\,,
\end{eqnarray}
where $P_\psi(\mathbf z)=|F_{\psi}(z)|^2\mu_\hbar(\mathbf z)$ (with $F_\psi$ the Segal-Bargmann transform of $|\psi\rangle$) has all features of a probability distribution, yet  in general $Q_{AW}^{-1}\left(Q_{Weyl}(A)\right)\neq A$. (For example, if $A(z)=x^2$, then $Q_{AW}^{-1}\left(Q_{Weyl}(A)\right)(z)=x^2-\hbar/2$.) It seems much more natural, if empirical adequacy permits, to assume that the expectation value is simply $\langle A\rangle_{|\psi\rangle} = \int A(\mathbf z)\,P_\psi(\mathbf z)\,d\mathbf z$, which leads one to conclude that Anti-Wick, not Weyl, quantization is the appropriate quantization scheme. To emphasize this important point once more: \textit{Only Anti-Wick quantization} gives exact equality between quantum expectation values and classical-looking phase-space averages over a positive distribution (the Husimi Q-function). No other standard quantization shares this property.

\section{Why the epistemic constraint?}

In Section 3, it was shown how expectation values calculated as phase space integrals can be rewritten as inner products of the familiar form Eq.\ (\ref{expectation_as_inner_product}) with self-adjoint operators obtained via Anti-Wick quantization \textit{if} probability distributions are of the form Eq.\ (\ref{SB_form}). If this is really the appropriate way of understanding quantum expectation values and, more broadly, the empirical success of quantum theory, this raises the question why turning to quantum theory effectively means restricting one's attention to probability distributions of the form Eq.\ (\ref{SB_form}) (and weighted sums thereof).

Before we sketch a possible, somewhat speculative, reply to this question, it should be highlighted that the restriction Eq.\ (\ref{SB_form}), which constrains the probability density to have the form of a Husimi Q-function, encodes a strong version of the Heisenberg uncertainty principle, interpreted as an epistemic, not ontic, constraint. One way of seeing this is by noting that, at any point $\mathbf z$ in phase space, the value of the Husimi function is constrained by \citep[p.\ 44]{hall_old}
\begin{eqnarray}
Q(\mathbf z)\le \frac{1}{(2\pi\hbar)^n}\,,
\end{eqnarray}
which imposes a limit on how concentrated the Husimi function can be in phase space.

To compare with Weyl quantization, where $x^2$ and $p^2$ are promoted to $\hat X^2$ and $\hat P^2$, respectively, we recall that the Heisenberg relations for the variances of $x$ and $p$, in Weyl quantization, are:
\begin{eqnarray}
(\Delta x)^2\,\cdot\,(\Delta p)^2 &=& \left(\langle x^2 \rangle-\langle x \rangle^2\right)\,\cdot\,\left(\langle p^2 \rangle-\langle p \rangle^2\right)\nonumber\\
&=& \left(\langle \hat X^2 \rangle-\langle \hat X \rangle^2\right)\,\cdot\,\left(\langle \hat P^2 \rangle-\langle \hat P \rangle^2\right)
 \ge \frac{\hbar}{4}. \label{Heisenberg}
\end{eqnarray}
In Anti-Wick quantization, $x^2$ and $p^2$ are promoted instead to
\begin{eqnarray}
Q_{AW}(x^2)&=&\hat X^2+\frac{1}{2}\hbar I\,,\nonumber\\
Q_{AW}(p^2)&=&\hat P^2+\frac{1}{2}\hbar I\,,  \label{AW_variances}
\end{eqnarray}
i.e. the occurrences of $\hat X^2$ and $\hat P^2$ in Eq.\ (\ref{Heisenberg}) are replaced by occurrences of $\hat X^2+\frac{1}{2}\hbar I$ and $\hat P^2+\frac{1}{2}\hbar I$, respectively, whereas the occurrences of $\hat X$ and $\hat P$ stay the same.

As a consequence, the product $(\Delta x)^2\,\cdot\,(\Delta p)^2$ of the variances is strictly larger in Anti-Wick quantization than in Weyl quantization (Eq.\ {\ref{Heisenberg}}), so the overall joint uncertainty -- which we interpret as a genuine, epistemic, uncertainty, about the true phase space location -- is actually larger than in Weyl quantization. This means that probability distributions which can be written in the form Eq.\ (\ref{SB_form}) have a minimal amount of combined ``spread'' in $x$- and $p$- directions that is larger than the combined spread of $x$ and $p$ in Weyl quantization.

What might be the reason that the switch from classical statistical mechanics to quantum theory should involve restricting attention to probability distributions with a nonzero minimal amount of spread, which play the role of Husimi Q-functions in the quantum theory, along the lines sketched in Section 4? As already alluded to in Section 2, this may be a consequence of hitherto not well understood microdynamics. Famously, the dynamics of quantum theories can be expressed using the Schr\"{o}dinger equation defined on Hilbert space vectors. For probability distributions $P(\mathbf z)$ of the form Eq.\ (\ref{SB_form}), which, as explained in Section 4, can be interpreted as Husimi Q-functions, such Schr\"{o}dinger-type evolution translates into a partial differential equation for the time evolution of the Q-function. \citet{drummond} has recently investigated such Q-function dynamics for a wide class of Hamiltonians, including ones in the Standard Model of elementary particles. He obtained the result that the time evolution equation for the Q-function has the form of a Fokker-Planck diffusion equation with a zero-trace diffusion matrix (see \citep[Sect. II]{drummond}). Solving this equation is not a well-posed problem if an initial probability density that is more concentrated than allowed by Eq.\ ({\ref{SB_form}}) are imposed, let alone a delta-function-shaped initial probability density, which would correspond to specifying the system's initial location $\mathbf z$ in phase space with arbitrary precision. (One may compare this to how a backwards-in-time heat equation is not solvable for arbitrarily sharp initial conditions \citep{miranker}.) \citet{drummond} also investigates which``microbehaviour'' would give rise to zero-trace Fokker-Planck time evolution at the level of the probability distribution and identifies a temporally bidirectional stochastic law of motion. Constraining the initial probability distribution so that it has the form of a Q-function, as is implicitly done by imposing Eq.\ (\ref{SB_form}), the initial value problem posed by the Fokker-Planck equation with zero trace becomes solvable.

We conclude that the restriction to probability distributions of the form Eq.\ (\ref{SB_form}) that was necessary when rewriting the expectation values of classical statistical mechanics as quantum theory obtained via Anti-Wick quantization presumably may be linked to solvability constraints on the time-evolution equations for $P(\mathbf z)$ (which in turn may be related to hitherto not well understood temporally bidirectional stochastic microdynamics). Consideration of those constraints can be bypassed by focusing on Hilbert space vectors and their time evolution, whether in Segal-Bargmann space or in some other, unitarily equivalent, Hilbert space. This may help to explain why quantum theory was developed in its Hilbert space form rather than as a version of classical statistical mechanics with a constraint on probability distributions

In any case, restricting one's attention to probability distributions of the form Eq.\ (\ref{SB_form}) should \textit{not} be understood as an \textit{ontic} constraint -- according to the view suggested here, the phase space location $\mathbf z$ is not in any way fuzzy but rather sharp and objective at any point in time -- only as an \textit{epistemic} constraint: $Q(\mathbf z)$ as an ``epistemic'' probability density expresses the assigning physicist's knowledge about the system's phase space location, but this knowledge cannot be arbitrarily concentrated (or ``peaked'') in phase space -- some spread of $Q(\mathbf z)$ is needed, otherwise its time evolution would not be well-posed.

\section{Conclusion and outlook}
Expectation values in quantum theory can be interpreted as integrals over phase space using a suitably smooth probability distribution -- but only if dynamical variables are mapped onto self-adjoint operators using Anti-Wick quantization rather than the more often used and more well-known Weyl quantization. The reading of quantum expectation values proposed here entails a reading of the Husimi Q-function as a proper probability function on phase space, a reading which seems to have first been considered by Bopp \citep{bopp} and recently ``rediscovered'' \citet{drummondreid,friederich}. It is worth noting that this point of view, while sharing with Bohmian mechanics the attraction that it assigns sharp values to all dynamical variables, has the additional interesting features of treating position and momentum on a par and of keeping intact the close connection between dynamical variables and self-adjoint operators, which is characteristic of standard quantum mechanics but absent in Bohmian mechanics. In that respect, one may see the account proposed here as ``less contextual'' than Bohmian mechanics.

Of course, empirical adequacy requires non-contextuality in the senses of Kochen-Specker and other no-go theorems \citep{spekkens,shrapnelcosta} to be violated. Kochen-Specker non-contextuality is the requirement that the values assigned to co-measurable dynamical variables must have the same algebraic relations as those variables themselves. In the present account, violation of this principle for the values assigned to self-adjoint linear operators is only to be expected -- and arguably well justified: In the present account the dynamical variables proper are simply the phase space functions $A(\mathbf q,\mathbf p)$. The self-adjoint linear operators $\hat A$ on which these functions are mapped by (Anti-Wick) quantization merely represent the functions for the purposes of calculations of expectation values and/or time evolution. But quantization procedure itself -- whether Anti-Wick or Weyl or Wick quantization -- does not preserve algebraic relations (see \citep{FriederichTyagi} for details), so Kochen-Specker non-contextuality was never a plausible assumption from the point of view of the present account in the first place. To assess how the present account relates to \emph{dynamical} criteria of non-contextuality such Spekkens' preparation non-contextuality and measurement non-contextuality \citep{spekkens}, we must take into account that, as shown in \citep[Sect.\ 7]{friederich}, any interpretation of the Q-function as an ``epistemic'' probability density -- i.e. a probability distribution that, unlike $|\psi|^2$ in Bohmian mechanics, is not an objective physical quantity and does not have a causal role -- must violate \emph{$\lambda$-mediation}, a key ingredient of  Spekkens' notions of non-contextuality (and, incidentally, of the assumptions from which the PBR theorem \citep{pbr} is derived). A discussion of whether the violation of $\lambda$-mediation is plausible will be given in a future publication together with a systematic account of the microdynamics that give rise to Q-function time evolution at the level of the probability density.

Finally, it is worth commenting briefly on the question of non-local correlations that violate Bell inequalities, which must of course be reproduced. This issue can only be addressed by considering the underlying microdynamics. As already emphasized by Drummond \citep{drummond} (and developed further in later work with Reid), the phase-space variables entering the Q-function must satisfy both forward- and backward-in-time oriented law-like probabilistic relations. The backward-oriented relations, referred to by Drummond and Reid as ``retrocausal'' \citep{drummondreid}, violate Bell's local causality and, hence, allow the violation of Bell inequalities. The mechanism through which this occurs is demonstrated for simple two- and three-mode systems in \citep{Friederich2025_EPL_EPR}. A systematic account of the microdynamics that implements these backward-oriented constraints will be provided elsewhere. Here I simply note that the issue of nonlocality is resolved at the dynamical level rather than at the level of Anti-Wick quantization itself.

The need for such microdynamics, already evident from the discussion of Bell correlations, becomes even more pressing when one turns to the measurement problem itself. Clearly such a systematic account is still very much needed: While the considerations presented here go some way towards solving the measurement problem by assigning sharp values to all dynamical variables, for a fully-fledged, completely satisfying solution to that problem a specification of a micro-law is clearly needed that, at the aggregate level, leads to Q-function dynamics. And while the results obtained by \citet{drummond} for bosonic quantum field theories go some way towards this goal, it would be important to see such dynamics written down explicitly as a generalization of Hamilton's equations, which describe the time evolution of the phase space variables in classical mechanics, preferably in a way that generalizes to relativistic and fermionic theories. Complementarily to this, it will be valuable to investigate the extent to which the considerations presented here can be generalized to complex manifolds beyond the phase space $\mathbb C^n$, where instead of Anti-Wick quantization one must consider the more general framework of Berezin (-Toeplitz) quantization (see \citep[Sect.\ 5]{ali} for a review) of which Anti-Wick quantization is a special case. The aim of this paper has been to show that Segal-Bargmann space and Anti-Wick quantization provide the natural structural starting point for such a programme.

\section*{Acknowledgements}
I would like to thank Jacob Barandes and two anonymous referees for useful comments on an earlier versions of this paper. I am grateful to Maurice de Gosson and Bryan Hall for help and Mritunjay Tyagi for fruitful discussions.

This research was funded by the Netherlands Organization for Scientific Research (NWO), project VI.Vidi.211.088.

\section*{Statement about AI use}
During the preparation of this work the author used OpenAI's o1- and GPT5-models for editing purposes. After using these tools, the author reviewed and edited the content as needed and takes full responsibility for the content of the published article.

\bibliography{antiWick_references}

\end{document}